\documentclass{article}

\usepackage{stefan_tex}

\usepackage{amsmath, amsthm, amssymb}
\usepackage{makecell}
\usepackage{multirow}
\usepackage{graphicx}
\usepackage{enumerate}
\usepackage{natbib}
\usepackage{url} 
\usepackage{caption}
\usepackage{subcaption}
\usepackage{fancyvrb}
\usepackage{enumerate}
\usepackage{relsize}
\usepackage{commath}
\usepackage{etoolbox}

\usepackage{hyperref}
\usepackage[margin=1.5in]{geometry}
\hypersetup{colorlinks,citecolor=blue,urlcolor=blue,linkcolor=blue}

\theoremstyle{plain}
\newtheorem{prop}{Proposition}

\theoremstyle{definition}

\theoremstyle{remark}

\begin{document}

\title{Sequential Validation of Treatment Heterogeneity}

\author{Stefan Wager \\ Stanford GSB}

\maketitle

\begin{abstract}
We use the martingale construction of \citet{luedtke2016statistical} to
develop tests for the presence of treatment heterogeneity. The resulting sequential
validation approach can be instantiated using various validation metrics, such as BLPs,
GATES, QINI curves, etc., and provides an alternative to cross-validation-like
cross-fold application of these metrics. This note was prepared as
a comment on the Fisher--Schultz paper by Chernozhukov, Demirer,
Duflo and Fern\'andez-Val, forthcoming in Econometrica.
\end{abstract}

\section{Introduction}

Consider a randomized controlled trial (RCT) on $i = 1, \, \ldots, \, n$ units with 
potential outcomes $Y_i(0), \, Y_i(1)$, pre-treatment covariates $Z_i$, and
a binary treatment $D_i \cond Y_i(0), \, Y_i(1), \, Z_i \sim \text{Bernoulli}(\pi)$.
The analyst observes $(Z_i, \, Y_i, \, D_i)$ where $Y_i = Y_i(D_i)$. As discussed
by in the paper by \citet*[henceforth CDDF]{chernozhukov2024fisher},
given data from an RCT it is often of interest to estimate and understand
treatment heterogeneity, and to use knowledge of such heterogeneity to design
treatment-assignment rules.

It is well understood that the conditional average treatment effect (CATE),
\begin{equation}
\tau(z) = \EE{Y_i(1) - Y_i(0) \cond Z_i = z},
\end{equation}
is a natural metric for quantifying treatment heterogeneity under the potential
outcomes model \citep{athey2016recursive,manski2004statistical}; furthermore, in an RCT,
the CATE admits a number
of simple identification formulas in terms of observables, e.g.,
$\tau(z) = \mathbb{E}[Y_i \cond Z_i = z, D_i = 1] - \mathbb{E}[Y_i \cond Z_i = z, D_i = 0]$.
What's less clear is how we should go about estimating the CATE. There's been
considerable recent interest in developing machine-learning based methods for
CATE estimation 
\citep[e.g.,][]{athey2019generalized,kunzel2019metalearners,hahn2020bayesian,foster2023orthogonal}.
However, consistency of these
methods relies on often hard-to-verify properties of the data-generating
distribution, and the same is true of the
fundamental statistical difficulty of the CATE estimation
task \citep{kennedy2022minimax}.

Given the challenges in obtaining provably valid estimates of $\tau(z)$ without
making regularity (e.g., smoothness) assumptions, CDDF recommend that applied
researches refrain from directly relying on theoretical consistency guarantees for non-parametric
CATE estimators. Instead, they recommend that researchers first make
a best effort to estimate $\htau(\cdot)$ well (e.g., using a machine learning
method), and then focus on evaluating how well the estimated $\htau(\cdot)$
captures the true treatment heterogeneity on held-out data. The main argument
of CDDF is that such evaluation metrics allow for simple and transparent
large-sample inference without making any smoothness assumptions on $\tau(\cdot)$.

I second this recommendation, and found CDDF to make a compelling case for it.
When estimating the CATE one of course wants
to try to be as accurate as possible. But then, when communicating results the transparency
and credibility of any claimed inferences is key---and validating a proposed $\htau(\cdot)$
estimate on a fresh evaluation dataset
is a much simpler task (and thus allows for more robust inferential strategies)
than than trying to estimate $\tau(\cdot)$ itself. The incentives are also well aligned:
If a researcher is able to start with a good estimate $\htau(\cdot)$, then such diagnostic
exercises will often yield interesting insights; whereas a researcher starting with a poor estimate
$\htau(\cdot)$ will simply find themselves unable to say anything at all about potential
treatment heterogeneity.

CDDF propose interrogating the alignment of $\htau(x)$ with the true $\tau(x)$ using
metrics which they refer to as the BLP and GATES.\footnote{The authors also discuss
the CLAN metric, but the focus of the CLAN is about describing the estimate $\htau(x)$
rather than validating its alignment with $\tau(x)$. The main use of CLAN comes after
we've already validated that $\htau(x)$ captures real and meaningful heterogeneity---thus
making a description of how this heterogeneity was captured relevant.}
The BLP metric estimates
the regression coefficient of the true CATE $\tau(Z_i)$ on estimates $\htau(Z_i)$
(and a coefficient of 1 means that $\htau(\cdot)$ is well calibrated); whereas
the GATES metric estimates average treatment effects across subgroups defined
by stratifying $\htau(Z_i)$ on the evaluation set (and verifies whether subgroups
with claimed larger treatment effects in fact do). Other relevant metrics that can
be used for this task include the QINI curve from the marketing literature 
\citep{radcliffe2007using},
the area-under-the-curve measure of \citet{zhao2013effectively},
and generalizations which \citet{yadlowsky2021evaluating} refer to as
rank-weighted average treatment effects (RATEs).

All these methods can be used to
produce test statistics that have expectation 0 when the estimates $\htau(Z_i)$ are
unaligned with any true treatment heterogeneity, and can be used to assess a
null of no (explained) treatment heterogeneity. CDDF construct confidence intervals
for both the BLP and GATES that hold under minimal assumptions (essentially the same assumptions needed
to build confidence intervals for the average treatment effect); large-sample inference
results for the QINI curve and its generalizations are available in \citet{yadlowsky2021evaluating}.

\section{Cross-Fold Validation}

One notable challenge with this approach---and a key focus of CDDF---is that
in many applications we seek to both provide an estimate $\htau(\cdot)$ and
evaluate it, and that splitting the data into separate training and evaluation
sets may be onerous. As such it is natural to explore cross-fold approaches.
The usual cross-validation-like blueprint for cross-fold evaluation is as follows:
\begin{enumerate}
\item Randomly split the data into $k = 1, \, \ldots, \, K$ evenly sized folds.
\item For each fold $k = 1, \, \ldots, \, K$, obtain an estimate $\htau^{(-k)}(\cdot)$ using the method
of one's choice applied to data from all but the $k$-th fold. 
\item Evaluate CATE estimates $\htau^{(-k)}(Z_i)$ for each observation $i$
belonging to the $k$-th fold.
\item Use these estimates, as well as the data from the $k$-th fold,
to produce a test statistics $T_k$ with the property that $T_k \Rightarrow \nn\p{0, \,1}$
under the null hypothesis of no (explained) treatment heterogeneity.
\end{enumerate}
One can leverage BLP, GATES, or other related metrics in step 4.

Now, given this setup, each of the $T_k$ on its own can obviously be used to
build a valid $p$-value against the null, $p_k = 2\Phi\p{-\abs{T_k}}$, where $\Phi(t)$
denotes the standard normal distribution. In practice, however, only using one of
the $T_k$ may feel wasteful, and so it may be of interest to provide $p$-values that
aggregate across all the $T_k$.

A first, na\"ive, approach to aggregation would be to treat the $K$ test statistics
as independent, and use
\begin{equation}
p_{naive} = 2\Phi\p{-\abs{\frac{1}{\sqrt{K}} \sum_{k = 1}^K T_k}}.
\end{equation}
This is, however, not valid, as the cross-fold construction doesn't imply independence
of the $T_k$; see the numerical experiment below for an example where these $p$-values
do not achieve nominal rejection rates. From a statistical perspective, the failures of
of na\"ive aggregation become acute when there is no treatment heterogneity to be found:
In this regime, the estimates $\htau^{(-k)}(\cdot)$ become highly responsive to the
noise in the training folds, thus creating dependence between the $T_k$.\footnote{The
usual argument deployed by \citet{chernozhukov2018double} to justify cross-fitting
does not hold here because these methods do not target a pathwise-differentiable parameter 
\citep{hirano2012impossibility}.}

What CDDF propose is to consider the median $p$-value instead:
\begin{equation}
p_{med} = \text{median}(p_k); \ \ \ \ p_k = 2\Phi\p{-\abs{T_k}}.
\end{equation}
Theorem 4.1 of CDDF establishes validity of this procedure under a median-concentration
condition.
The median aggregation approach is attractive in that it allows us to aggregate test
statistics across folds while maintaining validity. However, this approach still may
be seen as leaving information on the table. Each of the test statistics $T_k$ is
only computed using $n/K$ samples and so (especially when $K$ is large), any of
the $T_k$ on their own may not be particularly powerful. Ideally, one might hope
that aggregation would allow us to pool information across folds to recover power
comparable to a sample size of $n$ as occurs with, e.g., cross-fitting methods
for semiparametric inference \citep{chernozhukov2018double,schick1986asymptotically};
however, with median-aggregation this is not the case.

\citet{luedtke2016statistical} faced a similar challenge when studying inference for the value of the optimal
treatment rule, and proposed addressing it using a martingale construction that
goes through the folds in sequence and only uses data from past folds for estimation.
In our setting, this construction can be adapted into the following sequential validation procedure:
\begin{enumerate}
\item Randomly split the data into $k = 1, \, \ldots, \, K$ evenly sized folds.
\item For each fold $k = 2, \, \ldots, \, K$, obtain an estimate $\htau^{(1:(k-1))}(\cdot)$
using the method of one's choice applied to data from folds $1, \, \ldots, \, k-1$.
\item Evaluate estimates $\htau^{(1:(k-1))}(Z_i)$ for each observation $i$
belonging to the $k$-th fold.
\item As before.
\end{enumerate}
Finally, we aggregate the $p$-values as with the na\"ive construction:
\begin{equation}
p_{seq} = 2\Phi\p{-\abs{\frac{1}{\sqrt{K - 1}} \sum_{k = 2}^K T_k}}
\end{equation}
This sequential construction is able to pool information across folds but, unlike the
na\"ive method described above, is valid. The result below is
adapted from \citet{luedtke2016statistical}; the proof hinges on noting that,
under the assumptions made, the $T_k$ (approximately) form a martingale
difference sequence with predictable variance.

\begin{prop}
Suppose data from the $K$ folds are independent of each other and that, for each
$k = 2, \,, \ldots, \, K$, the test statistic
$T_k$ is valid under the null, i.e., $T_k \Rightarrow \nn\p{0, \, 1}$ conditionally on the provided
estimates $\htau^{(1:(k-1))}$. Then $p_{seq} \Rightarrow \operatorname{Unif}([0, \, 1])$ under the null.
\end{prop}

\section{A Numerical Experiment}

To better understand power of different methods for aggregating test statistics,
we consider a simple numerical evaluation on GATES with two-groups: Our test statistic
is the difference in ATEs for the top- vs bottom-half of units as ordered by
the estate $\htau(Z_i)$. We instantiate
the workflows described above as follows:
\begin{itemize}
\item We produce (training set) estimates $\htau(\cdot)$ using causal forests as implemented
in the \texttt{R} package \texttt{grf} of \citet{athey2019generalized}.
\item On the evaluation fold $k$, we produce groups $G_{k,1}$ and $G_{k,0}$ depending on
whether or not $\htau^{(-k)}(Z_i)$ (or $\htau^{(1:(k-1))}(Z_i)$) is greater than its
median value in that fold.
\item We produce our test statistic via an estimate of the treatment effect contrast,
\begin{equation}
\begin{split}
&\hdelta_k = \frac{\sum_{\cb{i : i \in G_{k,1}, \, D_i = 1}} Y_i}{\abs{i : i \in G_{k,1}, \, D_i = 1}} 
- \frac{\sum_{\cb{i : i \in G_{k,1}, \, D_i = 0}} Y_i}{\abs{i : i \in G_{k,1}, \, D_i = 0}} \\
&\ \ \ \ \ \ \ \ \ \ - \frac{\sum_{\cb{i : i \in G_{k,0}, \, D_i = 1}} Y_i}{\abs{i : i \in G_{k,0}, \, D_i = 1}} 
+ \frac{\sum_{\cb{i : i \in G_{k,0}, \, D_i = 0}} Y_i}{\abs{i : i \in G_{k,0}, \, D_i = 0}},
\end{split}
\end{equation}
along with a standard error estimate $\hsigma_k$ obtained via the jackknife \citep{efron1982jackknife};
we use the implementation in the \texttt{sandwich} package of \citet{zeileis2004econometric}.
The test statistic is $T_k = \hdelta_k / \hsigma_k$.
\item We use $K = 5$ folds.
\end{itemize}
We consider a simple RCT simulation with $n = 1,000$ units and $p = 10$ dimensional
covariates. We generate $Z_i \sim \text{Unif}([-1, \, 1]^p)$,
$Y_i(d) \cond Z_i \sim \nn\p{d\tau(z), \, 1}$, and $D_i \sim \text{Bernoulli}(\pi = 0.5)$.
We consider two settings for evaluation: One with $\tau(z) = 0$ (so the null holds),
and one with $\tau(z) = \p{z_1}_+$ (so treatment heterogeneity exists and is relatively
easy to find).

Results are shown in Table \ref{tab:res}. The first row shows rejection rates at a
nominal 5\% level under a setting where the null holds. We see that the na\"ive
cross-fold aggregation method fails, and gets a 10\% failure rate which is greater
than the nominal 5\% level. In contrast, both the median cross-fold aggregation and sequential validation
methods are valid (median aggregation is conservative, while sequential validation
is exact to within the precision of our simulation experiment).
The second row shows power in a setting where treatment heterogeneity
exists. Here, we see that the sequential validation method is much more powerful
than median aggregation. The reason for the improved power of sequential validation
is simple: For example, if all the $\hdelta_k$ are positive but none of the $T_k$
reaches significance on its own, then the sequential validation can take this shared
directionality into account whereas median aggregation cannot.

\begin{table}
\centering
\begin{tabular}{|c|ccc|}
\hline
rejection rate & na\"ive & median & sequential \\ \hline
$\tau(z) = 0$ & {\it 9.68\%} & 0.16\% & 5.21\%  \\
$\tau(z) = (z_1)_+$ & 83.40\%  &  21.14\%  & {\bf 65.52\%}  \\ \hline
\end{tabular}
\caption{Rejection probabilities for different $p$-value constructions,
with a 5\% significance level. In the first row, the null hypothesis holds
and so a rejection rate greater than 5\% implies that the method is not valid; we
denote this in italic. In the second row, we denote highest achieved rejection
rate (i.e., power) among valid methods in bold.
Results are aggregated across 10,000 simulation replications.}
\label{tab:res}
\end{table}

\section{Closing Remarks}

CDDF highlight the importance of robust and transparent validation tests
for treatment heterogeneity, and discuss a number of nice options. This
is a good recommendation, and I hope the CDDF paper will help promote the
use of such tests.
One seemingly esoteric and technical---yet in practice important---consideration is how to
develop in-sample validation tests that allow for both estimating $\htau(\cdot)$ and validating it
withing the same sample. Simple cross-fold aggregation, i.e., the na\"ive method,
is not valid here. CDDF propose median aggregation, which recovers validity at the cost
of lower power. Here, we argued that sequential validation appears more powerful than
median aggregation, all while retaining validity. Understanding optimal performance
for in-sample validation of treatment heterogeneity---and developing practical methods
that achieve such performance---seems like a promising goal for future work.

\section*{Acknowledgment}
I am grateful for many enlightening conversations with Susan Athey and Erik
Sverdrup on how best to estimate and validate heterogeneous treatment
effects in practice, and to Gene Katsevich for asking a very interesting
question. This research was supported by NSF grant SES--2242876.

\bibliographystyle{plainnat}
\bibliography{references.bib}

\end{document}